\documentclass[12pt]{article}
\pdfoutput=1
\usepackage[dvips]{graphicx}
\usepackage{epsfig, comment}
\usepackage{amsmath,amsthm,amsbsy,amssymb,amsfonts,graphicx,mathrsfs,bm,cite,color}
\usepackage{mathtools} 
\usepackage{blkarray}
\usepackage[all]{xy} 
\usepackage[makeroom]{cancel}
\usepackage[latin1]{inputenc}
\usepackage[american]{babel}
\usepackage[dvips]{graphicx}
\usepackage{color}
\usepackage[unicode]{hyperref}
\def\6{\partial}

\newcommand{\be}{\begin{equation}}
\newcommand{\ee}{\end{equation}}
\newcommand{\beq}{\begin{equation}}
\newcommand{\eeq}{\end{equation}}
\newcommand{\bea}{\begin{eqnarray}}
\newcommand{\eea}{\end{eqnarray}}
\newcommand{\nn}{\nonumber \\}

\newcommand{\ba}{\begin{eqnarray}}
\newcommand{\ea}{\end{eqnarray}}

\newcommand{\beqs}{\begin{eqnarray}}
\newcommand{\eeqs}{\end{eqnarray}}
\newcommand{\bal}{\begin{aligned}}
\newcommand{\eal}{\end{aligned}}

\def\lbldef#1#2{\expandafter\gdef\csname #1\endcsname {#2}}

\def\href#1#2{#2}

\newcommand{\ber}{\begin{eqnarray}}
\newcommand{\eer}{\end{eqnarray}}

\newcommand{\rme}{\operatorname{e}}
\newcommand{\beqar}{\begin{eqnarray}}

\newcommand{\eeqar}{\end{eqnarray}}

\newcommand{\dsl}
   {\kern.06em\hbox{\raise.15ex\hbox{$/$}\kern-.56em\hbox{$\partial$}}}

\newcommand{\eeqarr}{\end{eqnarray}}
\newcommand{\ZZ}{{\rm \kern 0.275em Z \kern -0.92em Z}\;}

\def\CC{{\mathchoice
{\rm C\mkern-8mu\vrule height1.45ex depth-.05ex
width.05em\mkern9mu\kern-.05em}
{\rm C\mkern-8mu\vrule height1.45ex depth-.05ex
width.05em\mkern9mu\kern-.05em}
{\rm C\mkern-8mu\vrule height1ex depth-.07ex
width.035em\mkern9mu\kern-.035em}
{\rm C\mkern-8mu\vrule height.65ex depth-.1ex
width.025em\mkern8mu\kern-.025em}}}

\def\RR{{\rm I\kern-1.6pt {\rm R}}}

\def\ZZ{{\rm Z}\kern-3.8pt {\rm Z} \kern2pt}
\def\IB{\relax{\rm I\kern-.18em B}}
\def\ID{\relax{\rm I\kern-.18em D}}
\def\II{\relax{\rm I\kern-.18em I}}
\def\IP{\relax{\rm I\kern-.18em P}}

\newcommand{\bear}{\begin{eqnarray}}
\newcommand{\eear}{\end{eqnarray}}

\def\to{\rightarrow}

\def\to{\rightarrow}

\def\6{\partial}

\newfont{\namefont}{cmr10}
\newfont{\addfont}{cmti7 scaled 1440}
\newfont{\boldmathfont}{cmbx10}
\newfont{\headfontb}{cmbx10 scaled 1728}

\newcommand{\dd}{\textrm{d}}

\newcommand{\cH}{\mathcal H}

\newcommand{\cO}{\mathcal O}

\newcommand{\rmT}{\text{T}}
\newcommand{\rmd}{{\mathrm{d}}}

\allowdisplaybreaks[1]

\numberwithin{equation}{section}

\begin{document}
\begin{titlepage}

\begin{flushright}
IPM/P-2018/076
\end{flushright}

\vskip 1.1cm
\begin{center}

\centerline{{\LARGE{\bf{Holographic integration of $T \bar{T}$ \& $J \bar{T}$ via $O(d,d)$}}}}

   \vskip 1.1cm

\centerline{{\large {\bf     T. Araujo$^a$, E. \'O Colg\'ain$^{a,b}$, Y. Sakatani$^c$, M. M. Sheikh-Jabbari$^d$, H. Yavartanoo$^e$}}}
       \vskip .6cm
             \begin{small}
               \textit{$^a$ Asia Pacific Center for Theoretical Physics, Postech, Pohang 37673, Korea}
               
               \vspace{3mm} 
               
               \textit{$^b$ Department of Physics, Postech, Pohang 37673, Korea}
               
               \vspace{3mm}
               
               \textit{$^c$ Department of Physics, Kyoto Prefectural University of Medicine, Kyoto 606-0823, Japan}
               
               \vspace{3mm}
               
               \textit{$^d$ School of Physics, Institute for Research in Fundamental Sciences (IPM),  \\ P.O.Box 19395-5531, Tehran, Iran}
               
               \vspace{3mm}
               
               \textit{$^e$ State Key Laboratory of Theoretical Physics, Institute of Theoretical Physics, \\ Chinese Academy of Sciences, Beijing 100190, China }
             \end{small}
\end{center}
\vskip 2mm
\begin{center} \textbf{Abstract}\end{center} \begin{quote}
Prompted by the recent developments in integrable single trace $T \bar{T}$ and $J \bar{T}$ deformations of 2d CFTs, we analyse such deformations in the context of AdS$_3$/CFT$_2$ from the dual string worldsheet CFT viewpoint. We observe that the finite form of these deformations can be recast as $O(d,d)$ transformations, which are an integrated form of the corresponding Exactly Marginal Deformations (EMD) in the worldsheet Wess-Zumino-Witten (WZW) model, thereby generalising the Yang-Baxter class that includes TsT. Furthermore, the equivalence between $O(d,d)$ transformations and marginal deformations of WZW models, proposed by Hassan \& Sen for Abelian chiral currents, can be extended to non-Abelian chiral currents to recover a well-known constraint on EMD in the worldsheet CFT. We also argue that such EMD are also solvable from the worldsheet theory viewpoint.

\end{quote} 

\vskip 3mm

\end{titlepage}
\setcounter{equation}{0}

\section{Introduction}
Recently a class of Lorentz invariant deformations of two-dimensional CFTs, dubbed $T \bar{T}$ deformations, have been demonstrated to be integrable \cite{Smirnov:2016lqw, Cavaglia:2016oda}. Concretely, one considers deforming the Lagrangian of a CFT by the perturbation $\delta \mathcal{L} = g T(z) \bar{T} (\bar{z})$, which is a composite operator of the holomorphic $T(z)$ and anti-holomorphic components $\bar{T}(\bar{z})$ of the stress-energy tensor. Since $T$ and $\bar{T}$ are operators of conformal dimension $(2,0)$ and $(0,2)$, respectively, the deformation is irrelevant and interpolates between an IR fixed point with conformal symmetry and a non-local UV QFT. See \cite{McGough:2016lol, Dubovsky:2017cnj, Shyam:2017znq, Guica:2017lia, Kraus:2018xrn, Cardy:2018sdv, Cottrell:2018skz, Bzowski:2018pcy, Aharony:2018vux, Bonelli:2018kik, Dubovsky:2018bmo, Taylor:2018xcy, Datta:2018thy, Conti:2018jho, Chen:2018eqk, Hartman:2018tkw, Aharony:2018bad, Aharony:2018ics, Cardy:2018jho, Conti:2018tca, Santilli:2018xux, Baggio:2018rpv,Chang:2018dge, Nakayama} for related work.

It is well-known that any consistent quantum theory of gravity in $AdS_3$ defines an asymptotic CFT \cite{AdS3-CFT2}. For this reason, the $T \bar{T}$ deformation is expected to have a holographic description. Two proposals exist: $T\bar T$ can be single or double trace. For double trace deformations, it has been argued that the deformation is dual to $AdS_3$ with a finite cutoff \cite{McGough:2016lol}. In contrast, the single trace counterparts give rise to deformed geometries \cite{Giveon:2017nie, Apolo:2018qpq, Chakraborty:2018vja} (see also \cite{Giveon:2017myj, Asrat:2017tzd, Giribet:2017imm, Baggio:2018gct, Chakraborty:2018kpr, Babaro:2018cmq, Chakraborty:2018aji}), which can be accessed through either their WZW, or alternatively supergravity description. While the deformations are irrelevant from the viewpoint of the holographic CFT at the $AdS_3$ boundary, they are marginal from the worldsheet CFT perspective, where using the general Kutasov-Seiberg construction \cite{AdS3-CFT2}, one can show that the worldsheet deformation is of $X\bar{Y}$-type where $X,Y$ are two worldsheet chiral currents. Through explicitly working out the spectrum (see references above), it is shown that the deformed worldsheet theory is solvable. The spectrum of the worldsheet theory matches that of the integrable dual CFT. In fact one can show that integrability or solvability of the deformed worldsheet theory follows from a similar analysis to that of $T{\bar T}$ \cite{Smirnov:2016lqw, Cavaglia:2016oda}, as the deformation has a similar form, i.e. $X{\bar Y}$-form \cite{Giveon:2017nie,Giveon:2017myj}.

Truly marginal deformations of two-dimensional CFTs are well studied. In particular, consider a deformation built from bilinear chiral currents obeying a current algebra, $\delta \mathcal{L} = g \sum_{ij} c^{ij} J_i(z) \bar{J}_j (\bar{z})$, where $c^{ij}$ are real constants and $g$ is a deformation parameter. As initially highlighted by Chaudhuri \& Schwartz (CS), a marginal deformation is guaranteed provided \cite{Chaudhuri:1988qb}
\be
\label{CYBE}
\sum_{i,j} c^{il} c^{jm} f_{ij}{}^{k} = \sum_{i,j} c^{li} c^{mj} \bar{f}_{ij}{}{}^{k} = 0, \quad \textrm{for all } l, m, k, 
\ee
where $f_{ij}{}^{k}$ and $\bar{f}_{ij}{}^{k}$ are the structure constants of the chiral current algebras. It is worth noting that this condition is effectively the Classical Yang-Baxter Equation (CYBE), where $c_{ij}$ is the $r$-matrix and the result applies directly to WZW or string $\sigma$-models. 

While \eqref{CYBE} guarantees marginality of the deformation to first order in $g$, i. e. perturbatively, writing down the Lagrangian for \textit{finite} $g$ may not be straightforward. This issue prompted Hassan \& Sen (HS) \cite{Hassan:1992gi} to conjecture that $O(d,d)$ transformations of any WZW model correspond to a marginal deformation of the WZW theory by an appropriate combination of Abelian chiral currents. This means that when $f_{ij}{}{}^{k}=\bar{f}_{ij}{}^{k}= 0$, so that the condition (\ref{CYBE}) is trivial, $O(d,d)$ presents us with arguably the most powerful means to identify the deformation at a finite value of the perturbation parameter. The proof of this conjecture is firmly established \cite{Kiritsis:1993ju, Henningson:1992rn}.  

In recent years the CYBE has also resurfaced in another context, notably the Yang-Baxter (YB) $\sigma$-model \cite{Klimcik:2002zj, Klimcik:2008eq}. The YB $\sigma$-model is specified by an $r$-matrix solution to the CYBE and presents one with a systematic way to deform a coset $\sigma$-model so that integrability is preserved. It can be applied to string $\sigma$-models to identify integrable deformations of $AdS_p \times S^p$ geometries \cite{Delduc:2013qra,Kawaguchi:2014qwa}. Subsequently, it was realised that the YB $\sigma$-model can be recast as a non-Abelian T-duality transformation \cite{Hoare:2016wsk, Borsato:2016pas, Borsato:2017qsx}, and in parallel that the YB $\sigma$-model constituted a generalisation of the (Seiberg-Witten) open-closed string map \cite{Seiberg:1999vs}, namely a single matrix inversion \cite{Araujo:2017jkb, Araujo:2017jap}. Regardless of which approach one adopts, one is led to the inevitable conclusion that the YB $\sigma$-model is a specific realisation of an $O(d,d)$ transformation in the target spacetime \cite{Sakamoto:2017cpu, Fernandez-Melgarejo:2017oyu, Sakamoto:2018krs, Lust:2018jsx}. In particular, for any supergravity solution with an isometry group, it was shown that there was a deformation based on the open-closed string map where the supergravity equations of motion reduce to the CYBE \cite{Bakhmatov:2017joy, Bakhmatov:2018apn}. The fact that this class of deformations exists can be explained from the perspective of non-Abelian T-duality \cite{Borsato:2018idb}.  Ultimately, these developments provide a concrete realisation of the ``gravity/CYBE correspondence" \cite{Matsumoto:2016lnr}.\footnote{See \cite{Araujo:2018rbc} for the extension to the modified CYBE and supergravity.}

Let us emphasise again that the HS proposal assumed the chiral currents were \textit{Abelian}. In the light of recent results in the YB literature, we now understand this restriction is unnecessary and it can be relaxed. Doing so, and provided we honour a ``unimodularity" condition \cite{Borsato:2016ose} that ensures the $O(d,d)$ transformation leads to a valid supergravity solution, so that the corresponding WZW model is still a CFT and the result (\ref{CYBE}) applies, we are expected to recover the condition (\ref{CYBE}) on the nose for any chiral current algebra. Admittedly, if one limits the scope to WZW models with groups $SL(2, \mathbb{R})$ or $SU(2)$, then even for rank four deformations that are unimodular \cite{Borsato:2016ose}, one is still restricted to chiral currents in the Cartan subalgebra, so there appear to be no examples where (\ref{CYBE}) is non-trivially satisfied. However, our arguments are more general and apply to any group beyond $SL(2, \mathbb{R})$ or $SU(2)$ where we expect non-Abelian deformations too. In summary, one of the main results of this paper is that (\ref{CYBE}) can be recovered from an $O(d,d)$ transformation. In this way, following \cite{Hassan:1992gi}, we provide a way to find finite transformations associated with the integration of the infinitesimal deformations. 

This leaves some comments on the single trace $T \bar{T}$ \cite{Giveon:2017nie} and $J \bar{T}$ deformations \cite{Apolo:2018qpq, Chakraborty:2018vja} in the literature. Despite being irrelevant from the perspective of the holographic CFT, they are marginal from the perspective of the string worldsheet CFT or WZW model, so they fall into the class of deformations discussed in \cite{Chaudhuri:1988qb}. Therefore, even in the case of non-Abelian symmetries, $O(d,d)$ provides a powerful method to identify the finite deformation, and as we show, it can be exploited to recover existing results and provide new examples. This replaces auxiliary Wakimoto variables \cite{Wakimoto:1986gf}, pursued as a means to identify the finite transformation in \cite{Giveon:2017nie}. In \cite{Apolo:2018qpq, Chakraborty:2018vja}, the infinitesimal deformation is exact, so the identification of the finite transformation was not a concern. We comment more on these examples in section \ref{sec:TsT}.

\textit{Note added}: While we were revising this manuscript, a related paper discussing overlapping material appeared on the arXiv \cite{Borsato:2018spz}. We explain the connection in the appendix.

\section{$T \bar{T}$ \& $J \bar{T}$: dual worldsheet viewpoint}
\label{sec:TsT}
In this section we review the known single trace current-current deformations, namely the $T{\bar T}$ and $J\bar T$. Let us begin with the deformation originally reported in \cite{Forste:1994wp} \footnote{See also \cite{Israel:2004vv, Detournay:2005fz} for related current-current deformations.}, which is later reproduced in \cite{Giveon:2017nie}, where auxiliary Wakimoto variables are evoked to find the finite transformation. As pointed out in \cite{Forste:1994wp}, while the infinitesimal transformation is the result of a deforming operator built from chiral currents, to get the finite transformation one can introduce gaugings in the spirit of the Buscher procedure \cite{Buscher:1987sk,Buscher:1987qj, Rocek:1991ps}. Here, we wish to spell out that the introduction and integration out of Wakimoto variables \cite{Wakimoto:1986gf} simply mirrors a TsT transformation \cite{Lunin:2005jy}, so that the final geometry is of course a TsT deformed geometry. 

A further message to the reader is that TsT with $B$-fields can be encapsulated in a single matrix inversion of the form \cite{Sakamoto:2018krs} (see also \cite{Borsato:2018idb}) 
\be
\label{open_closed}
[ (g+B)^{-1} + \Theta ]^{-1} = g' + B', \quad e^{-2 \Phi} \sqrt{-g} = e^{-2 \Phi'} \sqrt{-g'}, 
\ee
where $g, B$ are the original metric and $B$-field, $\Theta$ is the deformation parameter and $g', B'$ are the deformed metric and $B$-field. The dilaton transformation follows from a well-known T-duality invariant. When $\Theta$ is an antisymmetric product of Killing vectors, i. e. $\Theta^{m n} = -2 \eta r^{ij} v^{m}_i v^{n}_j$, where $r^{ij} = - r^{ji}$ are constants, the equations of motion of supergravity are expected \cite{Bakhmatov:2017joy} \footnote{When $B = 0$, we have both a perturbative \cite{Bakhmatov:2018apn} and a proof to all orders \cite{Bakhmatov:2018bvp}, while when $B \neq 0$, the map has been checked for a number of non-trivial cases \cite{Sakamoto:2018krs}. More generally, exploiting the connection to non-Abelian T-duality, this map can be derived \cite{Borsato:2018idb}.} to reduce to the CYBE. Specialising to the case where the skew-symmetric matrix $r^{ij}$ has only a single component with commuting Killing vectors, namely an Abelian $r$-matrix, we recover a simple TsT \cite{Osten:2016dvf}. One further comment is in order. When $B=0$ we recover the original open-closed string map of Seiberg-Witten \cite{Seiberg:1999vs}. We will come back to this map in the next section where we will review the known embedding of YB deformations in $O(d,d)$ \cite{Sakamoto:2018krs} with a view to extending it.

To illustrate YB deformations via the map (\ref{open_closed}), let us consider the $AdS_3 \times S^3 \times T^4$ geometry supported by $H$-flux: 
\bea
\label{ads3s3}
\dd s^2 &=& e^{2 \rho} ( - \dd t^2 + \dd x^2 ) + \dd \rho^2 + \frac{1}{4} ( \sigma_1^2 + \sigma_2^2 + \sigma_3^2) + \dd s^2(T^4), \nn 
H &=& - 2 e^{2 \rho} \dd t \wedge \dd x \wedge \dd \rho + \frac{1}{4} \sigma_1 \wedge \sigma_2 \wedge \sigma_3, 
\eea
where we have explicitly written out the $AdS_3$ components and adopted left-invariant one-forms $\sigma_i$ for the $S^3$, 
\be
\sigma_1 = - \sin \psi \dd \theta + \cos \psi \sin \theta \dd \phi, \quad \sigma_2 = \cos \psi \dd \theta + \sin \psi \sin \theta \dd \phi, \quad \sigma_3 = \dd \psi + \cos \theta \dd \phi. 
\ee
To get the finite transformation, one needs only specify the deformation parameter in the map (\ref{open_closed}), 
\be
\Theta = \frac{\eta}{2} \, \partial_t \wedge \partial_x, 
\ee
where $\eta$ is a deformation parameter. One slight subtlety is that the map can be singular in a particular gauge, since the matrix $G+B$ may be non-invertible. Provided one is careful to avoid this specific gauge, the map is well-defined and the end result of the deformation is \cite{Sakamoto:2018krs}: 
\bea
\label{TTdeformed}
\dd s^2 &=& \frac{e^{2 \rho} ( - \dd t^2 + \dd x^2 )}{(1 + \eta e^{2 \rho})} + \dd \rho^2 + \frac{1}{4} ( \sigma_1^2 + \sigma_2^2 + \sigma_3^2) + \dd s^2(T^4), \nn 
H &=& - \frac{2 e^{2 \rho} }{(1 + \eta e^{2 \rho})^2} \dd t \wedge \dd x \wedge \dd \rho + \frac{1}{4} \sigma_1 \wedge \sigma_2 \wedge \sigma_3. 
\eea
Note, this deformation corresponds to a TsT transformation in the $(t,x)$ directions. However, up to a Lorentz transformation it is the same as the deformation $\Theta = \eta \, \partial_{\gamma} \wedge \partial_{\bar{\gamma}}$, where we have introduced null coordinates, $\gamma = t + x, \bar{\gamma} = -t + x$. But therein lies another subtlety: T-duality, namely the ``T" of TsT is only well-defined for non-null directions. Up to some redefinitions, this is the same geometry discussed in \cite{Forste:1994wp, Giveon:2017nie}, so we conclude that it is a TsT geometry. 

However, one can go further with the comparison. We recall that Wakimoto variables \cite{Wakimoto:1986gf} were introduced in \cite{Giveon:2017nie}, essentially to identify the finite transformation, and in terms of these variables the worldsheet action may be expressed as (we set $\alpha_+ =1$)
\be
\label{action1}
\mathcal{L} = \partial \rho \bar{\partial} \rho - 2 R_{(2)} \rho + \beta \bar{\partial} \gamma + \bar{\beta} \partial \bar{\gamma} - \beta \bar{\beta} e^{-2 \rho}. 
\ee
The idea now is one integrates out the auxiliaries $\beta, \bar{\beta}$, which are quadratic in the action to recover the $AdS_3$ $\sigma$-model with coordinates $(\rho, \gamma, \bar{\gamma})$. The linear dilaton disappears in the procedure through the usual Gaussian path integral. 

Now, let us return to the original solution (\ref{ads3s3}) and T-dualise on the $x$-direction. Omitting the rest of the spacetime, which does not change, the resulting T-dual solution is
\bea
\dd s^2 = - 2 \dd x \dd t + e^{-2 \rho} \dd x^2 + \dd \rho^2, \quad \Phi = - \rho. 
\eea
It is worth noting that the $g_{tt}$ component of the metric has disappeared in the T-duality, an artifact of the fact that we considered the gauge where the combination $G+B$ is singular, and the geometry is supported solely by a linear dilaton without a $B$-field. Note, the matrix inversion (\ref{open_closed}) is singular in this gauge, but the TsT can be performed regardless. The corresponding worldsheet action is simply 
\be
\label{action_tdual}
\mathcal{L} = \partial \rho \bar{\partial} \rho + e^{- 2 \rho} \partial x \bar{\partial} x - \partial t \bar{\partial} x - \partial x \bar{\partial} t - 2 R_{(2)} \rho. 
\ee
To get back the original $\sigma$-model, one follows the Buscher procedure and gauges the isometry $x$, introducing a Lagrange multiplier $\tilde{x}$, integrating out the gauge fields, before finally adopting the gauge where the original coordinate $x$ is a constant. Let us rearrange the order here. First, we gauge the isometry direction $x$ by introducing the gauge fields, $\partial x \rightarrow \partial x + A, \bar{\partial} x \rightarrow \bar{\partial} x + \bar{A}$, before finally setting $x$ to be a constant. Following an integration by parts, the resulting action is 
\be
\mathcal{L} = \partial \rho \bar{\partial} \rho + e^{- 2 \rho} A \bar{A} - (\partial t + \partial \tilde{x}) \bar{A} + A (-\bar{\partial} t + \bar{\partial} \tilde{x}) - 2 R_{(2)} \rho. 
\ee
Note that up to the following redefintions:
\be
A \rightarrow \beta, \quad \bar{A} \rightarrow - \bar{\beta}, \quad (t + \tilde{x}) \rightarrow -\gamma , \quad (- t + \tilde{x}) \rightarrow \bar{\gamma}, 
\ee
this is nothing but the Giveon et al. action \cite{Giveon:2017nie, Giveon:2017myj} in Wakimoto variables (\ref{action1}). As noted in \cite{Giveon:2017nie}, the effect of the CFT deformation is to deform the worldsheet action by a term proportional to $\beta \bar{\beta}$, or in our notation $A \bar{A}$. To appreciate that this is the shift in TsT, let us return to the T-dual $\sigma$-model action (\ref{action_tdual}) and shift $ t \rightarrow t + \lambda x$ with $\lambda = - \frac{1}{2}$. Next, one repeats the gauging and redefines the variables to recover the deformed action, before finally integrating out the gauge fields. 

In summary, the holographic dual to the $T{\bar T}$ deformed 2d CFT is not only given by a TsT deformed geometry, but the introduction of auxiliary Wakimoto variables in the worldsheet action leads to an action that is equivalent to the T-dual $\sigma$-model action with gauged isometries. Integrating out these gauge fields, one completes the Buscher T-duality procedure. Thus, at every stage of the transformation one is mirroring a TsT transformation, and it is precisely this transformation that allows one to get a finite transformation. Finally, we note that while the conformal symmetry of the 2d CFT is broken (in the UV), the $T\bar T$ deformation keeps the Poincar\'e symmetry. This may be seen explicitly at the level of the dual gravity background and also the bulk worldsheet action. Moreover, this tells us that half the original target space supersymmetry is preserved by the deformation.

Since TsT transformations are Yang-Baxter deformations, essentially through the map (\ref{open_closed}), this raises the question are all current-current deformations in this class TsT transformations? To see that the answer to this is negative, let us consider the second example of a $J \bar{T}$ \cite{Chakraborty:2018vja} or $T \bar{J}$ deformation \cite{Apolo:2018qpq}. Here, we follow the treatment in \cite{Apolo:2018qpq}, but the deformation is the same modulo switching holomorphic and anti-holomorphic symmetries. The transformation differs from the earlier single trace $T \bar{T}$ transformation since one is mixing a current corresponding to the stress-energy tensor with a current from a global $U(1)$. In the concrete case we consider below, the $U(1)$ symmetry is a Cartan of an $SU(2)$ symmetry. 

Let us consider the original spacetime (\ref{ads3s3}), which we will now deform using our map (\ref{open_closed}) and the deformation parameter: 
\be
\label{gamma_psi}
\Theta^{\gamma \psi} = -\eta. 
\ee
Since $\gamma$ is a null direction, this transformation should be regarded as a T-duality on $\psi$ and a shift in $\gamma$, otherwise the T-duality does not make sense. The end result is 
\bea
\label{geom1}
\dd s^2 &=& \dd \rho^2 + e^{2 \rho} \left( \dd \gamma - \frac{\eta}{4} ( \dd \psi + 2 \cos \theta \dd \phi) \right)\dd \bar{\gamma} + \frac{1}{4} \left( \dd \theta^2 + \sin^2 \theta \dd \phi^2 + ( \dd \psi + \cos \theta \dd \phi)^2 \right), \nn
B &=& - \frac{1}{2} e^{2 \rho} \left( \dd \gamma - \frac{\eta}{4} ( \dd \psi + 2 \cos \theta \dd \phi) \right)\wedge \dd \bar{\gamma} + \frac{1}{4} \cos \theta \dd \phi \wedge \dd \psi, 
\eea
which up to signs and factors agrees with the result in \cite{Apolo:2018qpq}. To remove these factors, as explained in \cite{Azeyanagi:2012zd}, one can consider further shifts and rescaling in coordinates. Explicitly, one can redefine $\eta \rightarrow - 2 \eta$ and then shift $\gamma \rightarrow \gamma + \frac{\eta}{2} \psi$. The resulting geometry is \cite{Apolo:2018qpq}:
\bea
\label{TJdeformed}
\dd s^2 &=& \dd \rho^2 + e^{2 \rho} \left( \dd \gamma + \lambda ( \dd \psi + 2 \cos \theta \dd \phi) \right)\dd \bar{\gamma} + \frac{1}{4} \left( \dd \theta^2 + \sin^2 \theta \dd \phi^2 + ( \dd \psi + \cos \theta \dd \phi)^2 \right), \nn
B &=& - \frac{1}{2} e^{2 \rho} \left( \dd \gamma + \lambda ( \dd \psi + 2 \cos \theta \dd \phi) \right)\wedge \dd \bar{\gamma} + \frac{1}{4} \cos \theta \dd \phi \wedge \dd \psi. 
\eea
In summary, the finite form of the $T{\bar T}$ deformation is a TsT transformation, whereas the $J{\bar T}$ deformation requires an additional shift. This additional shift is evident in the $J{\bar T}$ example as the mapping (\ref{open_closed}) through the data (\ref{gamma_psi}) only executes a TsT transformation so that the intermediary solution is (\ref{geom1}). As explained above, to bring it to the eventual expression (\ref{TJdeformed}) an additional shift is required. Strictly speaking, this additional shift takes us outside the class of YB deformations, including TsT transformations, described by map (\ref{open_closed}) and  is more accurately viewed as an $O(d,d)$ transformation. One final comment: the end geometry preserves $\mathcal{N} = (2,0)$ supersymmetry, as we explain in the appendix.

\section{$O(d,d)$ transformations}
In the previous section we showed that there are deformations of the worldsheet CFT Lagrangian based on chiral currents that are not simple TsT transformations. They are also not YB deformations, but are best viewed as $O(d,d)$ transformations as conjectured initially in \cite{Hassan:1992gi}. Strictly speaking, there was the additional assumption that the chiral currents are Abelian, but in the light of the YB literature, this can be relaxed. In this section, we will solidify the connection. More concretely, we will show that given an undeformed non-linear $\sigma$-model with a Wess-Zumino term, $S_0$, and isometries generated by generalised Killing vectors, the result is a deformed $\sigma$-model $S_{\eta}$ expressed as a current-current deformation. Our treatment overlaps with \cite{Kiritsis:1993ju, Henningson:1992rn}. 

However, before proceeding to this result, let us review how TsT transformations and more generally YB deformations are embedded in $O(d,d)$ transformations. Consider a supergravity background that admits a set of Killing vectors, $v_i \equiv v_i^m \partial_m$, 
\be
\mathcal{L}_{v_i} g_{mn} = \mathcal{L}_{v_i} B_{mn} = \partial_{v_i} \Phi = 0. 
\ee
{The middle condition can be relaxed, as $v_i$ being an isometry of the full solution only requires ${\cal L}_{v_i} \dd B=0$, which is satisfied if 
\be
\mathcal{L}_{v_i} B =- \dd \tilde{v}_i,
\ee
for a one-form $\tilde{v}_i$. We can then introduce} the concept of a generalised Killing vector $(V_i^M)=(v_i^m,\,\tilde{v}_{im}\bigr)$. According to the introduction of the one-form $\tilde{v}_i$\,, a commutator of the generalised Killing vector is given by the Courant bracket \cite{Courant}, and we suppose $V_i$ form a closed algebra
\begin{align}
 [V_i,\,V_j]_C = f_{ij}{}^k\,V_k \,. 
\end{align}
We further suppose that they have constant inner products
\begin{align}
 {\cal G}_{ij} \equiv \eta_{MN}\,V_i^M\,V_j^N =v_i\cdot \tilde{v}_j+v_j\cdot \tilde{v}_i \,. 
\label{eq:calG}
\end{align}

In addition to these generalised Killing vectors, we introduce the standard notion of a generalised metric 
\begin{align}
 \cH_{MN} \equiv \begin{pmatrix} g_{mn} -B_{mp}\,g^{pq}\,B_{qn} & B_{mp}\,g^{pn} \\ -g^{mp}\,B_{pn} & g^{mn} \end{pmatrix}, 
\end{align}
and the T-duality invariant form of the dilaton: 
\begin{align}
 \rme^{-2d} \equiv \rme^{-2\Phi} \sqrt{-g}\,. 
\end{align}
We also introduce $\eta_{MN}$, a constant $O(d,d)$-invariant metric, 
\begin{align}
 \eta_{MN} \equiv \begin{pmatrix} 0 & \delta_m^n \\ \delta^m_n & 0 \end{pmatrix},
\end{align}
which allows us to raise and lower the $O(d,d)$ indices $M, N$. Having introduced some preliminaries, we are in a position to define the $O(d,d)$ transformation of interest to us. To do so, consider $n$ generalised Killing vector $V_i^{M}$, $i = 1, \dots, n$, and define the set of matrices: 
\begin{align}
 (T_{ij})_M{}^N \equiv V_{iM}\,V_j^N-V_{jM}\,V_i^N = - (T_{ji})_M{}^N \,.
\label{eq:Odd-generators}
\end{align}
It can be checked that the matrices satisfy the $o(d,d)$ property,
\begin{align}
 (T_{ij})_M{}^P\,\eta_{PN} + \eta_{MP}\,(T_{ij}^\rmT)^P{}_N = 0\,,
\end{align}
and as a result, they form a subalgebra of $o(d,d)$. Using these matrices, we define the relevant $O(d,d)$ matrix as
\begin{align}
\label{h}
 h = \rme^{-\eta \,r^{ij}\,T_{ij}}\,,\qquad r^{ij} = - r^{ji} \,, 
\end{align}
where $r^{ij}$ is a skew-symmetric matrix with constant entries, and the associated $O(d,d)$ transformation is
\begin{align}
 \cH'_{MN} = h_M{}^P\,\cH_{PQ}\,(h^\rmT)^Q{}_N \,,\qquad \rme^{-2d'}= \rme^{-2d}\,. 
\label{eq:Odd-trsf}
\end{align}
It is well known that the $O(d,d)$ transformation \eqref{eq:Odd-trsf} is a symmetry of string theory for constant $T_{ij}$. However, the generalised Killing vectors can depend on coordinates and in this case $T_{ij}$ is generically not a constant matrix. As a result, the $O(d,d)$ transformation is not guaranteed to give a supergravity solution. Nonetheless, it can be shown that there is a subgroup of $O(d,d)$, called $\beta$-transformations, where the transformation is equivalent to a YB deformation \cite{Sakamoto:2017cpu}. In this case, a solution is expected to exist once $r^{ij}$ is an $r$-matrix solution to the CYBE. See for example \cite{Bakhmatov:2017joy, Bakhmatov:2018apn}. 

Note, this transformation covers YB deformations and TsT as special cases, so they can all be repackaged as $O(d,d)$. Let us spell out how this happens. Truncating out the one-forms $\tilde{v}_i = 0$, the $O(d,d)$ matrix simplifies to 
\be
h = \left(\begin{array}{cc} \delta^{n}_{m} & 0 \\ \Theta^{mn} & \delta^m_n \end{array} \right), \quad \Theta^{mn} \equiv - 2 \eta r^{ij} v_i^m v_j^n, 
\label{eq:YB}
\ee 
and in the process the $O(d,d)$ transformation reduces to the map (\ref{open_closed}). Further specialising to the case where the Killing vectors commute, i. e. $[ v_i, v_j] = 0$, we recover a TsT transformation. This brings our definition of the $O(d,d)$ transformation to a close. We have explained the connection to YB deformations and all that is required is to introduce one-forms $\tilde{v}_i$ to describe more general transformations, such as the $T \bar{J}$ deformation, as we explain in the next section. 

\subsection{Finite worldsheet $J{\bar J}$ deformation as $O(d,d)$ transformations} Let us now return to the task of illustrating the connection between $O(d,d)$ transformations and current-current deformations of the string $\sigma$-model or WZW model. We start with the undeformed $\sigma$-model, 
\be
S_0 = \frac{1}{2 \pi} \int \dd^2z\, ( g_{mn} + B_{mn} )\, \bar{\partial} X^m \partial X^n. 
\ee
The Noether currents associated to the symmetries generated by the generalised Killing vectors are 
\begin{equation}\label{currents-general}
\begin{split}
 J_{(i)} &\equiv J_{(i)m}\,\partial X^m \equiv v_i^m\bigl(g+B)_{mn}\, \partial X^n + \tilde{v}_{in}\,\partial X^n \,, 
\\
 \bar{J}_{(i)} &\equiv \bar{J}_{(i)m}\,\bar{\partial} X^m \equiv v_i^m\,(g-B)_{mn}\, \bar{\partial} X^n - \tilde{v}_{in}\, \bar{\partial} X^n \,, 
\end{split}
\end{equation}
and they satisfy the equation $\bar{\partial} J_{(i)} + \partial \bar{J}_{(i)} = 0$. Let us now introduce the matrix 
\be
E_{mn} \equiv g_{mn} + B_{mn}, 
\ee
appearing in the $\sigma$-model action above. One can show under a generic $O(d,d)$ transformation, 
\be
h_M{}^N = \begin{pmatrix} \bm{p}_m{}^n & \bm{q}_{mn} \\ \bm{r}^{mn} & \bm{s}^m{}_n \end{pmatrix}, 
\label{Odd-h}
\ee
$E_{mn}$ transforms as 
\be
\label{transformedE}
E' = (\bm{q}+\bm{p}\,E)\,(\bm{s}+\bm{r}\,E)^{-1}. 
\ee
While this is true for an arbitrary $O(d,d)$ transformation, we can specialise to the case that interests us (\ref{h}), so that the $O(d,d)$ matrix can be expressed as 
\be
h_{M}^{~N} = \delta_{M}^{~N} + s^{ij}\, V_{iM}\,V_j^N
 = \begin{pmatrix} \delta_m^n+s^{ij}\,\tilde{v}_{im}\,v_j^n & s^{ij}\,\tilde{v}_{im}\,\tilde{v}_{jn} \\
 s^{ij}\,v_i^m\,v_j^n & \delta^m_n + s^{ij}\,v_i^m\,\tilde{v}_{jn}
\end{pmatrix}\,, 
\label{Odd-h-s}
\ee
where we have defined the following matrices: 
\begin{align}\label{s-calG}
 s^{ij} &\equiv -2\,\eta\,r^{ij} + \frac{(2\,\eta)^2}{2!}\,r^{ik}\,{\cal G}_{kl}\,r^{lj} - \frac{(2\,\eta)^3}{3!}\,r^{ik}\,{\cal G}_{kl}\,r^{lj}\,{\cal G}_{kl}\,r^{lj} + \cdots 
        = \sum_{n=1}^\infty \frac{(-2\,\eta)^{n}}{n!}\bigl[r\,({\cal G}\,r)^{n-1}\bigr]^{ij} \,, 
\end{align}
and the constant matrix ${\cal G}$ is defined in \eqref{eq:calG}. 
It is useful to note that this implies
\be\label{exp-eta-r}
\delta^i_j+s^{ik}{\cal G}_{kj}=(e^{-2\eta (r{\cal G})})^i{}_{j}. 
\ee
For the YB case with $\tilde{v}_i=0$, we have ${\cal G}_{ij}=0$ and hence $s^{ij}=-2r^{ij}$. 

In terms of these matrices, by comparing \eqref{Odd-h} with \eqref{Odd-h-s}, we can record the matrices appearing in the $O(d,d)$ transformation of the matrix $E$: 
\be
 (\bm{q}+\bm{p}\,E)_{mn} = E_{mn} + \,s^{ij}\,\tilde{v}_{im}\,J_{(j)n}, \quad (\bm{s}+\bm{r}\,E)^m{}_n = \delta^m_n + s^{ij}\,v_i^m\,J_{(j)n}. 
\ee
We now have to simply invert the latter, 
\be
 \bigl[(\bm{s}+\bm{r}\,E)^{-1}\bigr]_m{}^n = \delta_m^n - v_i^n\,s^{ik}\,t_k{}^j\,J_{(j)m} \,,
\ee
where
\be\label{t-lambda}
t_i{}^j \equiv \delta_i^j - J_{(i)m}\,v_k^m\,s^{kj} + \cdots = \frac{1}{\delta_i^j + J_{(i)m}\,v_k^m\,s^{kj}}\,. 
\ee
We next substitute these expressions back into (\ref{transformedE}) to find,
\be
 E'_{mn} = (E_{mp} +s^{ij}\,\tilde{v}_{im}\,J_{(j)p} )( \delta^p_n -\lambda^{ij} v_i^p\,J_{(j)n}) 
\ee
where 
\be\label{lambda-st}
 \lambda^{ij} \equiv s^{ik}\, t_{k}{}^{j}.
\ee
Next, from \eqref{t-lambda} we learn that
\be\label{lambda-s-X}
\lambda^{ij}=s^{ij}-s^{il} J_{(l)m}v_k^m \lambda^{kj}=s^{ij}-\lambda^{il} J_{(l)m}v_k^m s^{kj}, 
\ee
and hence
\bea 
E'_{mn}\equiv g'_{mn}+ B'_{mn} &=& E_{mn}-(\tilde{v}_{im}-E_{mp}v_i^p)\lambda^{ij} J_{(j)n}\nn
 &=& E_{mn} - \lambda^{ij} \, \bar{J}_{(i)m}\,{J}_{(j)n} \,.
 \label{eq:Eprime-lambda}
\eea
Therefore, the string $\sigma$-model action on the $O(d,d)$-transformed background may be expressed as:
\begin{align}
 S_\eta &= \frac{1}{2 \pi} \int \rmd^2z\, (g'_{mn}+B'_{mn})\partial X^m\bar\partial X^n \nonumber\\
 &= S_0 - \frac{1}{2 \pi} \int \rmd^2z\, \lambda^{ij}\, {\bar J}_{(i)}\,{J}_{(j)} \,.
\label{eq:deformed-action}
\end{align}
That is, the $O(d,d)$ transformed action can be interpreted as a current--current deformation of the original action. However, note that $\lambda^{ij}$ not only involves all powers of the deformation parameter $\eta$, but it also non-trivially depends on the worldsheet fields $X^m(z)$. We also bring the reader's attention to the fact that the action $S_\eta$ may be viewed as a one-parameter family of theories. Recalling \eqref{lambda-s-X}, 
\begin{align}
 \lambda^{ij} = 2\,\eta\,r^{ij} + \cO(\eta^2)\,,
\label{eq:action-series}
\end{align}
the infinitesimally deformed action is generated by $r^{ij}{\bar J}_{(i)} J_{(j)}$. As discussed in the introduction, this infinitesimal deformation is exactly marginal and also solvable if $r^{ij}$ satisfies the CYBE,
\begin{align}\label{CYBE-Doubled}
 f_{l_1l_2}{}^i\,r^{jl_1}\,r^{kl_2} + f_{l_1l_2}{}^j\,r^{kl_1}\,r^{il_2} + f_{l_1l_2}{}^k\,r^{il_1}\,r^{jl_2} =0\,, 
\end{align}
where $f_{ij}{}{}^k$ are the structure constants of the algebra of Killing vector fields. We note that \eqref{CYBE-Doubled} is a weaker condition than \eqref{CYBE} in the sense that it requires vanishing of the totally antisymmetric part of $r\cdot r\cdot f$ three-tensor. Note also that, unlike our case, for the CS deformations $c^{ij}$ is not necessarily anti-symmetric (recall that anti-symmetry of $r^{ij}$ is associated with the orthogonality of the generic $O(d,d)$ transformation \eqref{h}). 
We will elaborate on the CS condition \eqref{CYBE} and its relation to CYBE \eqref{CYBE-Doubled} in the next subsection. 

We now establish that $S_\eta$ may be viewed as integrated form of this infinitesimal deformation. To this end we begin by defining the ``transformed currents''
 \begin{equation}\label{transformed-currents-general}
\begin{split}
 J^\eta_{(i)} &\equiv J^\eta_{(i)m}\,\partial X^m \equiv v_i^m\bigl(g'+B')_{mn}\, \partial X^n + \tilde{v}_{in}\,\partial X^n \,, 
\\
 \bar{J}^\eta_{(i)} &\equiv \bar{J}^\eta_{(i)m}\,\bar{\partial} X^m \equiv v_i^m\,(g'-B')_{mn}\, \bar{\partial} X^n - \tilde{v}_{in}\, \bar{\partial} X^n \,,
\end{split}
\end{equation}
which can be expressed as follows by using matrices introduced above:\footnote{
These can be shown as follows. 
Starting with the definition \eqref{transformed-currents-general} and \eqref{eq:Eprime-lambda}, we obtain
\be
 J^\eta_{(i)}= J_{(i)}- v_i^m {\bar J}_{(k)m}\lambda^{kj} J_{(j)}=J_{(i)} -{J}_{(k)m}v_i^m \lambda^{kj} J_{(j)}+{\cal G}_{ik}\lambda^{kj} J_{(j)}=(1+{\cal G}s)_i{}^k t_k^j J_{(j)} ,\nonumber
\ee
where we used $ v_i^m {\bar J}_{(k)m}= {J}_{(k)m}v_i^m-{\cal G}_{ik}$, \eqref{lambda-st}, and \eqref{lambda-s-X}. 
Similarly one may work through ${\bar J}^\eta$:
\be
 \bar{J}^\eta_{(i)}=\bar{J}_{(i)}- v_i^m J_{(k)m} (\lambda^{\text T})^{kj}\bar{J}_{(j)}= (s^{-{\text T}} \lambda^{\text T})_i{}^j {\bar J}_{(j)}.\nonumber
\ee
where we used \eqref{lambda-st} and \eqref{lambda-s-X}.}
\be\label{J-eta-J}
 J^\eta_{(i)}= (\delta_i^k +{\cal G}_{il} s^{lk}) t_k{}^j J_{(j)},\qquad
 {\bar J}_{(i)}^\eta =(s^{-{\text T}} \lambda^{\text T})_i{}^j {\bar J}_{(j)} .
\ee
Then, we can show the following identity for the transformed action $S_\eta$:
\begin{subequations}
\begin{align}
 S_{\eta+\delta\eta} &= S_\eta - \frac{\delta\eta}{2\pi} \int \dd^2z\, \frac{\rmd \lambda^{ij}}{\rmd \eta} {\bar J}_{(i)} J_{(j)} \label{S-eta-deformed-1} \\ 
 &\equiv S_\eta - \frac{\delta\eta}{2\pi} \int \dd^2z\, C^{ij} \, \bar{J}^\eta_{(i)}\, {J}^\eta_{(j)} \,, \label{S-eta-deformed-3}
\end{align}
\end{subequations}
where as we will show below
\begin{align}
C^{ij}=-2r^{ij}.
\end{align}

To establish the above, let us start with $\frac{\rmd \lambda^{ij}}{\rmd \eta} {\bar J}_{(i)} J_{(j)}=C^{ij} \, \bar{J}^\eta_{(i)}\, {J}^\eta_{(j)}$. Employing the identity,
\begin{align}
 \frac{\rmd\lambda^{ij}}{\rmd \eta}= (\lambda\,s^{-1})^i{}_k\,\frac{\rmd s^{kl}}{\rmd \eta} t_l{}^j , 
\end{align}
which can be shown from \eqref{lambda-s-X}, we obtain, 
\begin{align}
C^{ij} \, \bar{J}^\eta_{(i)}\, {J}^\eta_{(j)} &= \frac{\rmd s^{kl}}{\rmd \eta} (s^{-{\text T}} \lambda^{\text T})_k{}^i {\bar J}_{(i)} t_l{}^j J_{(j)} \nonumber\\ 
 &= \frac{\rmd s^{ik}}{\rmd \eta}[(1+{\cal G}s)^{-1}]_k{}^j {\bar J}_{(i)}^\eta J^\eta_{(j)},
\end{align}
where \eqref{J-eta-J} is used. 
This leads to
\begin{align}
 C^{ij}&= \frac{\rmd s^{ik}}{\rmd \eta}[(1+{\cal G}s)^{-1}]_k{}^j = -2r^{ij},
\end{align}
where in the last equality we used \eqref{exp-eta-r}. 
This completes the derivation of our formula
\begin{align}
 S_{\eta+\delta\eta} = S_\eta - \frac{\delta\eta}{\pi} \int \dd^2z\, r^{ij} \, {J}^\eta_{(i)} \, \bar{J}^\eta_{(j)}\,. 
\label{S-eta-deformed-2}
\end{align}

We discuss the physical interpretation in the next subsection.

\subsection{Physical implications and discussions}

As we already discussed, infinitesimal deformations \eqref{eq:action-series} are exactly marginal if $r^{ij}$ satisfies the CYBE \eqref{CYBE-Doubled}. 
On the other hand we have the CS condition \eqref{CYBE} which as discussed in \cite{Chaudhuri:1988qb} guarantees marginality of the $g \sum_{ij}\ c^{ij}J_{(i)} {\bar J}_{(j)} $ deformations to all orders in $g$. On a different account, as discussed, the action at finite deformation parameter $\eta$ has three important features, (1) $S_{\eta}$ can be written as a deformation by the same currents ${\bar J}_{(i)}, {J}_{(j)}$ as in \eqref{eq:deformed-action} and (2) $S_{\eta+\delta\eta}-S_{\eta}$ can be written as a deformation by $J_{(i)} {\bar J}_{(j)}$ with a \emph{field dependent} coupling $\rmd \lambda^{ij}/\rmd\eta$ \eqref{S-eta-deformed-1} or (3) $S_{\eta+\delta\eta}-S_{\eta}$ can be written as a deformation with transformed currents $J^\eta_{(i)} {\bar J}^\eta_{(j)}$ with a \emph{constant} coupling  $-2r^{ij}$, as given in \eqref{S-eta-deformed-2}. In this part we would like to elaborate on these results.

\paragraph{More on CS condition \eqref{CYBE} vs. CYBE \eqref{CYBE-Doubled}.}
In the setup studied in \cite{Chaudhuri:1988qb}, existence of the holomorphic and anti-holomorphic currents are assumed. 
Correspondingly, we prepare a set of the left and right generalised Killing vectors, $\{V_i\}=\{V_a,\,\bar{V}_{\bar{b}}\}$ satisfying
\begin{align}
 [V_a,\, V_b]_{C} = f_{ab}{}^c\,V_c \,,\quad 
 [\bar{V}_{\bar{a}},\, \bar{V}_{\bar{b}}]_{C} = \bar{f}_{\bar{a}\bar{b}}{}^{\bar{c}}\,\bar{V}_{\bar{c}} \,, 
\end{align}
which commute with each other. 
The left and right symmetries are characterised by\footnote{Note that, if we denote $(V_a^M)=(v_a^m,\,\tilde{v}_{am})$ and $(\bar{V}_{\bar{a}}^M)=(v_{\bar{a}}^m,\,\tilde{v}_{\bar{a}m})$, these equations are equivalent to $v_a^n\,(g_{nm}-B_{nm}) - \tilde{v}_{an} = 0$ and $v_{\bar{a}}^n\,(g_{nm}+B_{nm}) + \tilde{v}_{\bar{a}n} = 0$, which make the associated Noether currents holomorphic or anti-holomorphic.}
\begin{align}
 \cH^M{}_N\,V_a^N = + V_a^M\,,\qquad \cH^M{}_N\,\bar{V}_{\bar{a}}^N = - \bar{V}_{\bar{a}}^M\,. 
\end{align}
Under the setup, the associated Noether currents $\{J_i\}=\{J_a,\,\bar{J}_{\bar{b}}\}$ are (anti-)holomorphic. 

Due to the left-right split, the $O(d,d)$ generators are decomposed as $\{T_{ij}\}=\{T_{ab},\,T_{a\bar{b}},\,T_{\bar{a}\bar{b}}\}$\,. 
We can easily see that an $O(d)\times O(d)$ subgroup generated by $T_{ab}$ and $T_{\bar{a}\bar{b}}$ does not deform the supergravity background. 
Therefore, the deformations are essentially parameterized by $O(d,d)/O(d)\times O(d)$, namely $h = \rme^{-2\eta \,r^{a\bar{b}}\,T_{a\bar{b}}}$\, \footnote{At first sight the situation here appears opposite to the usual formulation of $O(d,d)$ transformations, where the non-trivial transformations are contained in the $O(d) \times O(d)$ subgroup. However, it should be noted that here we are deforming the geometry by introducing a bivector corresponding to a Drinfeld twist and the non-trivial deformations are of the form $r^{a \bar{b}}$.}. 
For this type of $r$-matrix, which is unimodular, the CYBE reduces to
\begin{align}
 r^{d\bar{a}}\,r^{e\bar{b}}\,f_{de}{}^{c} = 0\,,\quad 
 r^{a\bar{d}}\,r^{b\bar{e}}\,\bar{f}_{\bar{d}\bar{e}}{}^{\bar{c}} = 0\,,
\label{eq:marginal}
\end{align}
which is exactly \eqref{CYBE} by identifying $c_{ij}$ and $r^{a\bar{b}}$\,. 

\paragraph{On $\lambda^{ij}J_{(i)} {\bar J}_{(j)}$ deformation and its exact marginality.} 
To argue for marginality of finite deformation we first {show} that the coupling $\lambda^{ij}$ satisfies \eqref{CYBE} provided that $r^{ij}$ satisfies the same equation for both $f_{ij}{}{}^k, {\bar f}_{ij}{}{}^k$ structure constants. To see this it is enough to note that 
$$ s^{ij}=r^{ik} A_{k}{}^j=B^i{}_k r^{kj}$$
for some known matrices $A, B$. This may be readily seen from \eqref{s-calG}. Then recalling \eqref{lambda-s-X} this implies 
\be\label{lambda-X-Y}
\lambda^{ij}=r^{ik} X_{k}{}^j=Y^i{}_k r^{kj}
\ee
for some known matrices $X,Y$. However, the CS condition \eqref{CYBE} only guarantees exact marginality of the deformation for constant $c^{ij}$. In our case $\lambda^{ij}$ is not a constant and has a non-trivial field dependence. Nonetheless, we note that the dependence of coupling $\lambda^{ij}$ on worldsheet fields appears through $X^n(z,\bar z)$ from components of generalized Killing vector $v_i, \tilde{v}_i$ and the background fields $g,B$ through $J_{(i)}^m$. One can then see that the analysis of \cite{Chaudhuri:1988qb} goes through and 
the correlators involving any number of the deformation term still vanish upon assuming \eqref{CYBE}.

\paragraph{Integrated deformations as $O(d,d)$ transformations.} 
As constructed and is explicitly and nicely demonstrated in \eqref{eq:deformed-action} and \eqref{S-eta-deformed-2}, our $O(d,d)$ transformed action produces a one-parameter family of theories and  perturbation around any point of them produces the same form of the deformation. That is the $O(d,d)$ transformation provides a way to integrate the infinitesimal YB-deformation generated by $r^{ij}J_{(i)} {\bar J}_{(j)}$, as discussed in \cite{Hassan:1992gi,Henningson:1992rn}. While the CS deformation does not ask for antisymmetry of the coupling, our analysis suggests that the condition of being able to integrate the deformation to a finite transformation requires its antisymmetry, so that there is an associated $O(d,d)$ transformation.\footnote{We would like to remark that one should distinguish between exact marginality of the deformation and the condition of its integrability; while the latter requires the former, the exact marginality does not imply existence of associate finite transformation.}

As another crucial remark we note that given a background with generalized Killing vectors $V_i^M$, these vectors do not remain Killing under a generic $O(d,d)$ transformation. Therefore, the transformed currents $J^\eta_{(i)}, {\bar J}^\eta_{(i)}$ \eqref{transformed-currents-general} are \emph{not} generically conserved anymore.
In the case of Abelian $r$-matrix, where $[V_i,\,V_j]_C=0$, we can easily show that $V_i$ satisfy the generalised Killing equations even in the finitely transformed background. 
Therefore, the ``transformed currents'' $J^\eta_{(i)}$ and $\bar{J}^\eta_{(i)}$ are conserved Noether currents. 
Then, the current-current deformation \eqref{S-eta-deformed-2} from $S_\eta$ to $S_{\eta+\delta\eta}$ is exactly marginal because the coupling $r^{ij}$ is a solution of CYBE. 
In this sense, the finite $O(d,d)$ transformation yields an ``integrated'' version of the deformation. 

Note that this general result extends the analysis of \cite{Hassan:1992gi,Henningson:1992rn} to general $O(d,d)$ deformations of arbitrary backgrounds admitting Abelian generalised Killing vectors.
In non-Abelian cases, where $V_i$ do not commute with each other, the formula \eqref{S-eta-deformed-2} is still valid but a part of the isometries generated by $V_i$ can be broken in the deformed geometry, and $J^\eta_{(i)}$ are not necessarily conserved. This makes the interpretation of \eqref{S-eta-deformed-2} less clear. 
However, as we know from many examples, an $O(d,d)$ transformed geometry is always a solution of supergravity if the $r$-matrix satisfies the CYBE. 
This indicates that the deformed theory $S_\eta$ for a finite $\eta$ is CFT, and the deformation $S_\eta \to S_{\eta+\delta\eta}$ should be a marginal deformation.

\section{Examples}
Having established that single trace $T \bar{T}$ and $T \bar{J}$ deformations are more accurately regarded as $O(d,d)$ transformed solutions, we can turn our attention to revisiting the existing examples in this language. Along the way, we will take the opportunity to illustrate some new related examples. For concreteness, we consider deformations of the geometry $AdS_3 \times S^3 \times T^4$ (\ref{ads3s3}). Our conventions and notations are gathered in appendix \ref{Appendix-A}. Inevitably, this will restrict us to pretty trivial examples \footnote{This is even true if one extends the scope beyond rank two examples to rank four deformations obeying the unimodularity condition \cite{Borsato:2016ose}, so that the worldsheet theory is still a CFT.} with commuting chiral currents, but the generality of the results in the previous section should not be overlooked. For more general WZW models, they are expected to apply. 

\paragraph{$T{\bar T}$ case.} 
To begin, let us recall the $O(d,d)$ transformation of interest \eqref{eq:Odd-trsf} and to get oriented, let us rehash the original $T \bar{T}$ deformation \cite{Forste:1994wp} from this more general vantage point. To do so, one simply has to identify the pair of generalised Killing vectors, 
\be
 V_1 = (\partial_{\gamma},0), \quad V_2 = (\partial_{\bar{\gamma}}, 0). 
\ee
Note, in the absence of the additional one-forms $\tilde{v}$, this deformation reduces to a YB deformation with $r$-matrix $r = \frac{1}{2} \partial_{\gamma} \wedge \partial_{\bar{\gamma}}$. More specifically, this transformation is a TsT transformation, and exploiting Lorentz symmetry, it is equivalent to a TsT transformation in the $(t,x)$-directions, which is well-defined from the perspective of T-duality, as we explained in section \ref{sec:TsT}. In this case, the deformed $\sigma$-model action is 
\be
S_{\eta} = S_0 - \frac{\eta}{2 \pi} \int \dd^2 z \rme^{2 \Phi'} J^- \bar{J}^- = S_0 - \frac{\eta}{2 \pi} \int \dd^2 z \frac{1}{(1 + \eta \rme^{2 \rho})} J^- \bar{J}^-. 
\ee
where $\Phi'$ denotes the deformed dilaton and $J^- = \rme^{2 \rho} \partial \bar{\gamma}$, $\bar{J}^- = \rme^{2 \rho} \bar{\partial} \gamma$ correspond to chiral currents. Modulo a coordinate change, the geometry can be found in (\ref{TTdeformed}). 

\paragraph{$J{\bar T}$ case.} 
As stated, the previous example is a vanilla YB deformation (or TsT). The $J \bar{T}$ deformation, since it requires an additional shift, is not in this class and we presently turn our attention to it. The additional shift necessitates adding dual one-forms. So, here we define the generalised Killing vectors: 
\be
V_1 = (\partial_{\gamma}, 0), \quad V_2 = ( \partial_{\psi}, - \frac{1}{4} \dd \psi). 
\ee
What is interesting about this example is the $O(\eta^2)$ terms do not appear in the final expression and the infinitesimal transformation agrees with the finite one. In the end the deformed $\sigma$-model may be expressed as 
\be
S_\eta = S_0 - \frac{\eta}{2 \pi} \int \dd^2 z\, J^-\, \bar{k}^3, 
\ee
where we have introduced the current $\bar{k}^3 = \frac{1}{2} ( \bar{\partial} \psi + \cos \theta \bar{\partial} \phi)$. Before departing this example, let us remark that in the absence of the one-form, we recover the deformed geometry (\ref{geom1}), which is a TsT deformed geometry, as explained earlier. 

\paragraph{$T{\bar T}$ and $J{\bar T}$ combined.} 
To do something new, we can easily combine the previous two examples by considering
\begin{align}
 V_1 = (\partial_\gamma,\,0)\,,\qquad V_2 =\bigl(c_1\,\partial_{\bar{\gamma}}+c_2\,\partial_\psi,\, - \frac{c_2}{4} \rmd\psi \bigr)\,,
\end{align}
where we can truncate to the $T\bar{T}$ or $J\bar{T}$ deformation, respectively, by choosing $c_2=0$ or $c_1=0$\,. 
The infinitesimal deformation of the $\sigma$-model action is obviously marginal,
\begin{align}
 S_\eta &= S_0 - \frac{\eta}{2\pi} \int \rmd^2z\, \frac{1}{(1+c_1\,\eta \rme^{2\rho})}\, J_- \, \bigl(c_1\,\bar{J}_- +c_2\,\bar{k}_3\bigr) 
\nn
 &= S_0 - \frac{\eta}{2\pi} \int \rmd^2z\, \bigl(c_1\,J_- \,\bar{J}_- +c_2\,J_- \,\bar{k}_3\bigr) + \cO(\eta^2)\,, 
\end{align}
where in the second line we have expanded to isolate the infinitesimal deformation. 

\paragraph{$K{\bar T}$ deformation.} 
Let us consider one further example, which is related to the $T \bar{T}$ deformation, but the anti-holomorphic $SL(2, \mathbb{R})$ of Kutasov-Seiberg \cite{AdS3-CFT2} has been subjected to an $SL(2,\mathbb{R})$ transformation. We know that this deformation exists, as in the usual notation of conformal symmetry generators where $P$ denote translation and $K$ special conformal generators, specialised to null coordinates, the $r$-matrix $r = \frac{1}{2} P_{\gamma} \wedge K_{\gamma}$ is a trivial solution to the CYBE. In terms of our current framework, the deformation is specified by the following combination of generalised Killing vectors: 
\begin{align}
 V_1 = (\partial_\gamma,\,0)\,,\qquad 
 V_2 = \bigl(-\bar{\gamma}\,\partial_\rho -\rme^{-2\rho} \partial_{\gamma}+\bar{\gamma}^2\,\partial_{\bar{\gamma}},\,\bar{\gamma}\,\rmd\rho+\rmd\bar{\gamma}\bigr)\,,
\end{align}
and the $\sigma$-model action in the deformed geometry becomes
\begin{align}
 S_\eta &= S_0 - \frac{\eta}{2\pi} \int \rmd^2z\,\frac{1}{(1 + \eta\rme^{2\rho}\bar{\gamma}^2)}\, J_- \, \bar{J}_+ 
\nn
 &= S_0 - \frac{\eta}{2\pi}\int \rmd^2z\,J_- \, \bar{J}_+ + \cO(\eta^2)\,,
\end{align}
where $\bar{J}_+ = -2\,\bar{\gamma}\,\bar{\partial}\rho +\rme^{2\rho}\bar{\gamma}^2\,\bar{\partial}\gamma -\bar{\partial}\bar{\gamma}$\,. As with the previous example, it is hard to imagine that one can identify the fully integrated deformation to all orders in $\eta$ by exploiting Wakimoto variables and this makes $O(d,d)$ transformations potentially the only game in town.

\section{Discussion} 
This manuscript began life as a question: are single trace $T \bar{T}$ \cite{Giveon:2017nie} and $J\bar{T}$ \cite{Apolo:2018qpq, Chakraborty:2018vja} deformations TsT or YB transformations? As explained in section \ref{sec:TsT}, this question can be answered in the negative. A related puzzle concerned the role of the Wakimoto variables, which appear in the $T \bar{T}$ deformation as a means to extend the infinitesimal current-current deformation to a finite geometric transformation, whereas for the $J \bar{T}$ deformation, it turns out they are unnecessary as the infinitesimal deformation is exact. Funnily enough, the answer to these questions were already in the literature and just needed to be resurrected.

\begin{figure}[t]
    \centering
    \includegraphics[width=0.8\linewidth]{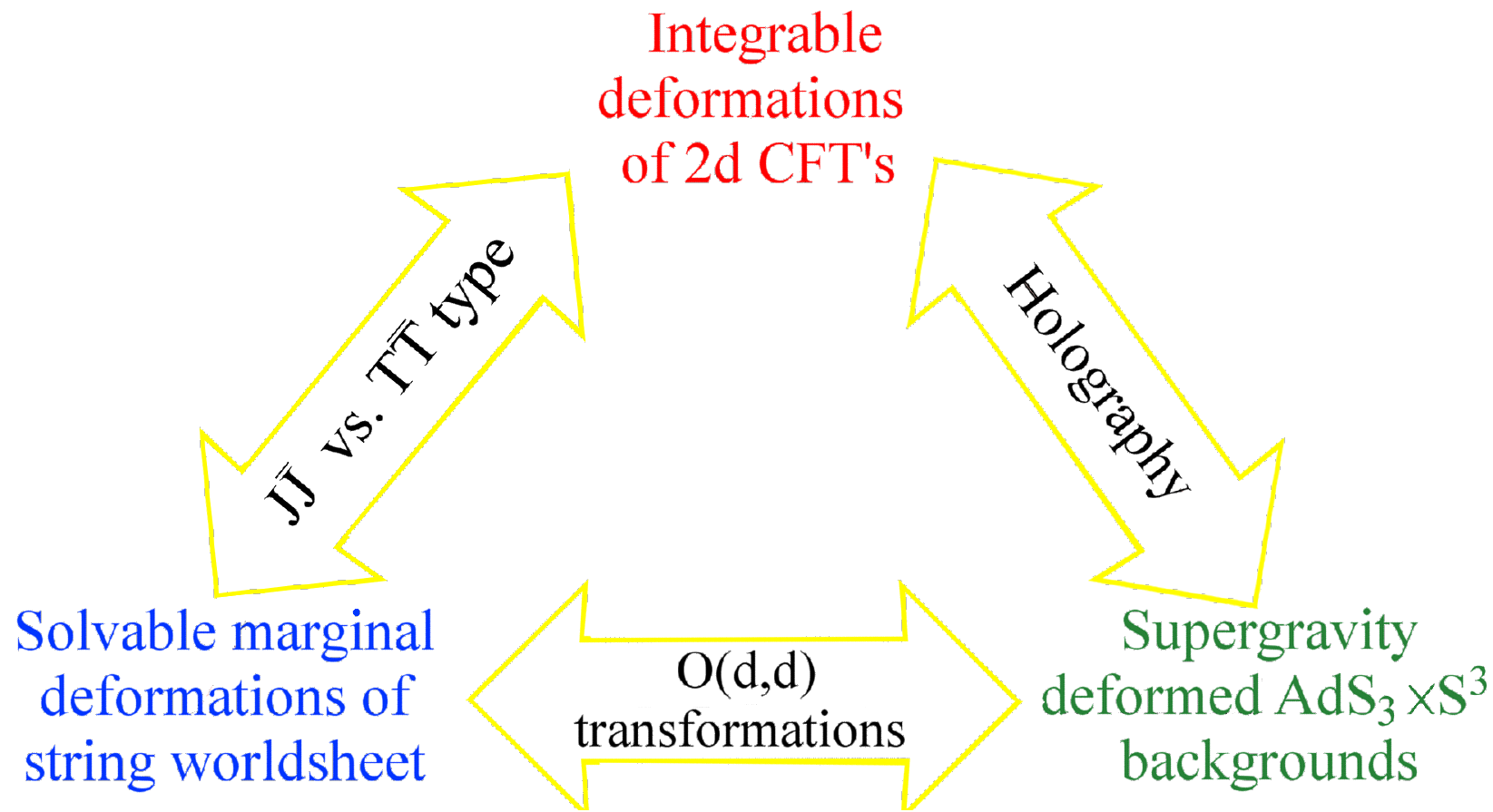}
    \caption{Triangle of integrable deformations. The top corner denotes the integrable, irrelevant deformations of a 2d CFT, like single trace $T{\bar T}$ or $J{\bar T}$ deformations. The bottom left corner shows the general $J{\bar J}$ deformations, which are marginal deformations of the string worldsheet theory on $AdS_3\times S^3$ background. These marginal deformations are dual to the irrelevant deformations of the 2d CFT. It has been shown in the literature that these deformations of WZW model are also solvable. As we argued here $O(d,d)$ transformations allow us to find finite deformations in this class. These finite deformations can then be traced back in the 2d CFT side. The bottom right corner shows the gravity background obtained from the same $O(d,d)$ transformation. We should mention here that the string theory WZW model have another class of integrable deformations, the YB deformations that are dual to ``noncommutative'' 2d CFT. These deformations, too, are generated by another class of $O(d,d)$ transformations, which in especial cases reduce to TsT.}
 \label{Figure}
\end{figure}

To appreciate this, one simply has to recall that $T \bar{T}$ and $J \bar{T}$ deformations, while irrelevant from the perspective of the holographic CFT are in fact marginal for the worldsheet CFT. As such, these deformations fall into the CS class with chiral currents. However, they are indeed more restricted and fit into the CS class with \textit{Abelian} chiral currents, a setting where the condition (\ref{CYBE}) is in fact trivial. This is precisely the context where $O(d,d)$ transformations were proposed \cite{Hassan:1992gi} as a powerful method to identify the integrated transformation corresponding to a given infinitesimal marginal current-current deformation. 

Our contribution here is importing insights from the YB literature and in particular the understanding of the YB deformation as an $O(d,d)$ transformation. This allowed us to generalise the HS prescription to $O(d,d)$ transformations based on non-constant matrices, thereby recovering the CS result, which is valid for non-Abelian current algebras, on the nose. This is new. We have also grasped the opportunity to illustrate how the Noether currents transform under an $O(d, d)$ transformation. 

With this new perspective, we revisited the single trace $T\bar{T}$ and $J \bar{T}$ deformations and illustrated how they may be understood within $O(d,d)$. This provides a powerful alternative to Wakimoto variables that is expected to work in all cases. We have also showcased the connection to generalised Killing vectors and the role of appropriately chosen one-forms. Building on this technique we constructed a number of further examples of related deformations that are not accessible to a Wakimoto description. 

In the examples such as $T \bar{T}$ and $J\bar{T}$ deformations, the deformation of the geometry was always realised as a combination of a TsT transformation and an additional coordinate transformation. 
Here we briefly explain the reason for that. 
In these examples, the generalised Killing vectors satisfy $r^{ij}\,\tilde{v}_i\wedge\tilde{v}_j=0$ and this makes the $O(d,d)$ matrix to the form
\begin{align}
 h = \rme^{-\eta \,r^{ij}\,T_{ij}} = \begin{pmatrix} a & 0 \\ b & a^{-\rmT} \end{pmatrix} = \begin{pmatrix} a & 0 \\ 0 & a^{-\rmT} \end{pmatrix} \begin{pmatrix} 1 & 0 \\ \Theta & 1 \end{pmatrix},\qquad \Theta \equiv a^\rmT b\,.
\end{align}
The first matrix on the right hand side represents a diffeomorphism $GL(d)$ while the second matrix is the YB-type deformation \eqref{eq:YB}. 
In other words, the $O(d,d)$ transformation can be replaced by a combination of a YB-type deformation and a coordinate transformation. 
In particular, when $V_i^M$ are constant like the case of $T \bar{T}$ and $J\bar{T}$ deformations, this is exactly a combination of TsT transformation and a shift.

Our discussion here was mainly focussed on the $ AdS_3\times S^3$ background and its dual 2d CFT. However, in our string worldsheet or WZW theory analysis we did not use this specific background in any crucial way. This analysis can hence be readily generalized to WZW models associated with coset spaces other than $AdS_3\times S^3$, in particular to $AdS_4\times \mathbb{C} P_3$ or $AdS_5\times S^5$ backgrounds. The analogue of $T{\bar T}$ deformation in the 3 or 4 dimensional CFTs dual to these backgrounds has been briefly mentioned in \cite{Hartman:2018tkw, Taylor:2018xcy}. Nonetheless, as argued our analysis here is dual to a single trace deformation in the 3d or 4d CFTs and the deformations discussed in these papers are double trace. As in the 2d case, one may construct such single trace deformations starting from an orbifold CFT. We hope to explore this direction in future publications.

\section*{Acknowledgements} 
We would like to thank I. Bakhmatov, E. Musaev, J. Sakamoto and K. Yoshida for comments and discussions. The work of YS is supported by the JSPS KAKENHI Grant Number JP 18K13540 and 18H01214. The work of MMShJ is supported in part by the Saramadan-Iran grant no. ISEF/M/97219 and INSF young scientist research chair grant no 950124. The work of HY is supported in part by National Natural Science Foundation of China, Project 11675244. This research was also facilitated by the Ministry of Science, ICT \& Future Planning, Gyeongsangbuk-do and Pohang City. 

\appendix

\section{Comments on the relation to \cite{Borsato:2018spz}}

After submitting our original manuscript to the arXiv, a preprint \cite{Borsato:2018spz} appeared, which performed a more detailed analysis of deformations of $AdS_3\times S^3$. 
In our original manuscript, when we made the connection to the CS condition, we mainly focussed on the ``strong CS condition'' in the terminology of Borsato \& Wulff \cite{Borsato:2018spz}. This condition is the necessary and sufficient condition for marginality of the deformation for $\sigma$-models with compact groups. For the non-compact groups, however, the condition for marginality of the deformation, as was originally noted in the CS paper \cite{Chaudhuri:1988qb}, is a bit weaker and has been dubbed ``the weak CS condtion'' \cite{Borsato:2018spz}. In the class of $AdS_3\times S^3$ deformations, there are cases where the weak CS condition is satisfied while not the strong one. Here, we explain the difference and clarify our original claim. 

Let us consider a background which has a set of left and right generalised Killing vectors, $\{J_i\}=\{J_a,\,\bar{J}_{\bar{a}}\}$. 
Since a general group manifold will have such a set of left/right generalised Killing vectors, this assumption is not so restrictive. 
On the other hand, in order to derive the CS condition, we further supposed that our $O(d,d)$-transformation is generated only by $T_{a\bar{b}}$\,, or in other words, we have assumed $r^{ab}=0$ and $r^{\bar{a}\bar{b}}=0$\,. 
This is more restrictive as it excludes non-unimodular $r$-matrices. 
We then showed that the CYBE for the unimodular $r$-matrix $r^{a\bar{b}}$ is precisely the CS condition \eqref{eq:marginal} (which is called the strong CS condition in \cite{Borsato:2018spz}). 
Therefore, unlike the comment made in \cite{Borsato:2018spz} on our work,  we do not claim the equivalence of the CYBE based on a general $r$-matrix and the strong CS condition.

\paragraph{Examples beyond the strong CS condition.}

Here, going beyond our original restriction, we show through examples that for a non-unimodular $r$-matrix the strong CS condition is broken. Consider a non-unimodular $r$-matrix, $r=\frac{1}{2}\,V_1\wedge V_2$ with
\begin{align}
\begin{split}
 V_1 &= c_1\, V_- + c_2\, \bar{V}_- = (c_1\,\partial_\gamma+c_2\,\partial_{\bar{\gamma}},\,0)\,,
\\
 V_2 &= V_3+ \bar{V}_3 = \bigl(- \partial_\rho + \gamma\,\partial_\gamma + \bar{\gamma}\,\partial_{\bar{\gamma}},\, 0 \bigr)\,, 
\end{split}
\label{eq:non-unimodular}
\end{align}
which was studied in section 4.2.1 of \cite{Sakamoto:2018krs}. 
If the product of the constants is zero, $c_1\,c_2 =0$, the deformed supergravity background satisfies the supergravity equations of motion. Otherwise, it is a solution to the generalised supergravity field equations \cite{Arutyunov:2015mqj, Wulff:2016tju}.

As we have discussed, in the case of a group manifold, we can always decompose an $O(d,d)$ matrix into a product of $O(d,d)/O(d)\times O(d)$ and an automorphism $O(d)\times O(d)$\,. 
In the present example, we can decompose the $O(d,d)$ matrix as $h_M{}^N = (\rme^{h_1} \rme^{h_0})_M{}^N$ with
\begin{align}
\begin{split}
 &h_1 \equiv - \eta\,\bigl( c_1 \,T_{-\bar{3}} - c_2 \,T_{3\bar{-}} + \tfrac{\eta\,c_1\,c_2}{2}\,T_{-\bar{-}} \bigr) \,,
\\
 &h_0 \equiv - \eta\,\bigl( c_1 \,T_{-3} + c_2 \,T_{\bar{-}\bar{3}} \bigr) \,.
\end{split}
\end{align}
Since the automorphism $\rme^{h_0}$ does not deform the background, the non-unimodular deformation \eqref{eq:non-unimodular} is equivalent to the deformation associated with a non-Abelian $r$-matrix,
\begin{align}
 r'= \tfrac{c_1}{2}\,V_-\wedge \bar{V}_3 - \tfrac{c_2}{2}\,V_3\wedge \bar{V}_- + \tfrac{\eta\,c_1\,c_2}{4}\,V_-\wedge \bar{V}_- \,.
\label{eq:r-matrix-CS}
\end{align}

One should note that  although $\rme^{h_0}$ does not have any effect on the deformation, in fact, it is playing the role of relaxing the CYBE. 
Originally, the $r$-matrix $r=\frac{1}{2}\,V_1\wedge V_2$ satisfies the CYBE for arbitrary $c_1$ and $c_2$, but after modding out $\rme^{h_0}$, the $r$-matrix \eqref{eq:r-matrix-CS} satisfies the CYBE (or the strong CS condition) only when $c_1\,c_2 =0$\,. 
In our setup, since we have assumed $r^{ab}=0$ and $r^{\bar{a}\bar{b}}=0$\,, we have not considered the case $c_1\,c_2 \neq 0$\,, where the strong CS condition is broken. 
On the other hand, the $r$-matrix \eqref{eq:r-matrix-CS} satisfies the weak CS condition of \cite{Borsato:2018spz}, which is the condition for the exact marginality and is equivalent to the strong CS condition only for compact groups. 
This means that the conformal invariance of the string $\sigma$-model is realised even for solutions of generalised supergravity.

Other $r$-matrices, called $\mathbf{R_1}$\,, $\mathbf{R_4}$\,, and $\mathbf{R_9}$, which violate the strong CS condition were discussed in \cite{Borsato:2018spz}. 
Although these $r$-matrices originally satisfy the CYBE, just as in the above example, after decomposing the $O(d,d)$-matrices as $h = \rme^{-2\eta\,r'^{a\bar{b}}\,T_{a\bar{b}}} \rme^{-\eta\,(a^{ab}\,T_{ab}+b^{\bar{a}\bar{b}}\,T_{\bar{a}\bar{b}})}$ and ignoring the second factor, the $r$-matrix $r'^{a\bar{b}}$ does not satisfy the CYBE. 
Therefore, this is again beyond our assumption, where $r'^{a\bar{b}}$ was assumed to follow the CYBE. 

It is also interesting to note that in some examples a non-unimodular $r$-matrix can become unimodular after decomposing the $O(d,d)$-matrix as $h = \rme^{-2\eta\,r'^{a\bar{b}}\,T_{a\bar{b}}} \rme^{-\eta\,(a^{ab}\,T_{ab}+b^{\bar{a}\bar{b}}\,T_{\bar{a}\bar{b}})}$ and ignoring the second factor. 
Indeed, in the above example $r=\frac{1}{2}\,V_1\wedge V_2$ with \eqref{eq:non-unimodular}\,, if we choose $c_2=0$\,, this $r$-matrix is non-unimodular because $r^{ab}\,f_{ab}{}^c\neq 0$\,. 
However, after the factorization $h = \rme^{h_1} \rme^{h_0}$ and the ignoring the second factor, the $r$-matrix \eqref{eq:r-matrix-CS} with $c_2=0$ is Abelian. 
This clearly explains the reason why some non-unimodular $r$-matrices can give solutions to conventional (and not generalised) supergravity \cite{Sakamoto:2018krs,Wulff:2018aku}. 

\section{Generalised Killing vectors in $AdS_3\times S^3$}\label{Appendix-A}

In the background \eqref{ads3s3}, we find the following set of generalised Killing vectors:
\begin{align}
 V_- &\equiv(\partial_\gamma,\,0)\,,\qquad 
 V_3 \equiv\bigl(-\tfrac{1}{2}\,\partial_\rho +\gamma\,\partial_\gamma,\,- \tfrac{\rmd\rho}{2}\bigr)\,,
\\
 V_+&\equiv\bigl(-\gamma\,\partial_\rho +\gamma^2\,\partial_{\gamma}-\rme^{-2\rho} \partial_{\bar{\gamma}},\, - \gamma\,\rmd\rho - \rmd\gamma \bigr)\,,
\\
 \bar{V}_- &\equiv(\partial_{\bar{\gamma}},\,0)\,,\qquad 
 \bar{V}_3 \equiv \bigl(-\tfrac{1}{2}\,\partial_\rho +\bar{\gamma}\,\partial_{\bar{\gamma}},\,\tfrac{\rmd\rho}{2}\bigr)\,,
\\
 \bar{V}_+&\equiv\bigl(-\bar{\gamma}\,\partial_\rho -\rme^{-2\rho} \partial_{\gamma}+\bar{\gamma}^2\,\partial_{\bar{\gamma}},\, \bar{\gamma}\,\rmd\rho+\rmd\bar{\gamma} \bigr)\,,
\\
 K_1 &\equiv \Bigl(\sin\phi\,\partial_\theta+\tfrac{\cos\phi}{\tan\theta}\,\partial_\phi-\tfrac{\cos\phi}{\sin\theta}\,\partial_\psi,\, \tfrac{\sin\phi\,\rmd\theta-\tfrac{\cos\phi}{\tan\theta}\,\rmd\phi-\tfrac{\cos\phi}{\sin\theta}\,\rmd\psi}{4}\Bigr)\,,
\\
 K_2 &\equiv \Bigl(\cos\phi\,\partial_\theta-\tfrac{\sin\phi}{\tan\theta}\,\partial_\phi+\tfrac{\sin\phi}{\sin\theta}\,\partial_\psi,\, \tfrac{\cos\phi\,\rmd\theta+\tfrac{\sin\phi}{\tan\theta}\,\rmd\phi+\tfrac{\sin\phi}{\sin\theta}\,\rmd\psi}{4}\Bigr)\,,
\\
 \bar{K}_1 &\equiv \Bigl(\sin\psi\,\partial_\theta-\tfrac{\cos\psi}{\sin\theta}\,\partial_\phi+\tfrac{\cos\psi}{\tan\theta}\,\partial_\psi,\, \tfrac{-\sin\psi\,\rmd\theta+\tfrac{\cos\psi}{\sin\theta}\,\rmd\phi+\tfrac{\cos\psi}{\tan\theta}\,\rmd\psi}{4}\Bigr)\,,
\\
 \bar{K}_2 &\equiv \Bigl(\cos\psi\,\partial_\theta+\tfrac{\sin\psi}{\sin\theta}\,\partial_\phi-\tfrac{\sin\psi}{\tan\theta}\,\partial_\psi,\,- \tfrac{\cos\psi\,\rmd\theta+\tfrac{\sin\psi}{\sin\theta}\,\rmd\phi+\tfrac{\sin\psi}{\tan\theta}\,\rmd\psi}{4}\Bigr)\,,
\\
 K_3 &\equiv \bigl(\partial_\phi,\, \tfrac{\rmd\phi}{4}\bigr)\,, \qquad 
 \bar{K}_3 \equiv \bigl(\partial_\psi,\,- \tfrac{\rmd\psi}{4}\bigr)\,. 
\end{align}
These satisfy the algebra
\begin{align}
\begin{split}
 [V_3,\,V_\pm]_C &= \pm V_\pm\,,\quad 
 [V_+,\,V_-]_C = -2\,V_3\,,\quad
 [\bar{V}_3,\,\bar{V}_\pm]_C = \pm \bar{V}_\pm\,,\quad 
 [\bar{V}_+,\,\bar{V}_-]_C = -2\,\bar{V}_3\,,
\\
 [K_1,\,K_2]_C &= K_3\,,\quad 
 [K_2,\,K_3]_C = K_1\,,\quad 
 [K_3,\,K_1]_C = K_2\,, 
\\
 [\bar{K}_1,\,\bar{K}_2]_C &= \bar{K}_3\,,\quad 
 [\bar{K}_2,\,\bar{K}_3]_C = \bar{K}_1\,,\quad 
 [\bar{K}_3,\,\bar{K}_1]_C = \bar{K}_2\,,
\end{split}
\end{align}
and the corresponding Noether currents are holomorphic or anti-holomorphic, having the following form:
\begin{align}
 J_- &= \rme^{2\rho}\partial\bar{\gamma} \,, \quad
 J_3= -\partial\rho + \rme^{2\rho}\gamma\,\partial\bar{\gamma} \,, \quad
 J_+= -2\,\gamma\,\partial\rho -\partial\gamma+\rme^{2\rho}\gamma^2\,\partial\bar{\gamma} \,,
\\
 \bar{J}_- &= \rme^{2\rho}\bar{\partial}\gamma \,, \quad
 \bar{J}_3 = -\bar{\partial}\rho+\rme^{2\rho}\bar{\gamma}\,\bar{\partial}\gamma \,, \quad
 \bar{J}_+ = -2\,\bar{\gamma}\,\bar{\partial}\rho +\rme^{2\rho}\bar{\gamma}^2\,\bar{\partial}\gamma -\bar{\partial}\bar{\gamma} \,,
\\
 k_1 &= \frac{1}{2}\,\bigl(\sin\phi\,\partial\theta - \sin\theta\,\cos\phi\,\partial\psi\bigr)\,, \quad
 k_2 = \frac{1}{2}\,\bigl(\cos\phi\,\partial\theta + \sin\theta\,\sin\phi\,\partial\psi\bigr)\,,
\\
 \bar{k}_1 &= \frac{1}{2}\,\bigl(\sin\psi\,\bar{\partial}\theta - \sin\theta\,\cos\psi\,\bar{\partial}\phi\bigr)\,, \quad
 \bar{k}_2 = \frac{1}{2}\,\bigl(\cos\psi\,\bar{\partial}\theta + \sin\theta\,\sin\psi\,\bar{\partial}\phi\bigr)\,,
\\
 k_3 &= \frac{1}{2}\,\bigl(\partial\phi + \cos\theta\,\partial\psi\bigr)\,, \quad
 \bar{k}_3 = \frac{1}{2}\,\bigl(\bar{\partial}\psi + \cos\theta\,\bar{\partial}\phi \bigr)\,.
\end{align}
For the sets of left and right generalised Killing vectors
\begin{align}
 \{V_a\}\equiv \{V_+,\,V_-,\,V_3,\,K_1,\,K_2,\,K_3\}\,,\qquad 
 \{\bar{V}_{\bar{a}}\}\equiv \{\bar{V}_+,\,\bar{V}_-,\,\bar{V}_3,\,\bar{K}_1,\,\bar{K}_2,\,\bar{K}_3\}\,,
\end{align}
the constant matrices ${\cal G}_{ab}$ and $\bar{{\cal G}}_{\bar{a}\bar{b}}$ become
\begin{align}
 &({\cal G}_{ab}) = \frac{1}{2} \left(\begin{smallmatrix}
 0 & -2 & 0 & 0 & 0 & 0 \\
 -2 & 0 & 0 & 0 & 0 & 0 \\
 0 & 0 & 1 & 0 & 0 & 0 \\
 0 & 0 & 0 & 1 & 0 & 0 \\
 0 & 0 & 0 & 0 & 1 & 0 \\
 0 & 0 & 0 & 0 & 0 & 1 
 \end{smallmatrix}\right) = -(\bar{{\cal G}}_{\bar{a}\bar{b}}) \,.
\end{align}

\section{Supersymmetry of $T \bar{J}$ deformation}
Let us solve the Killing spinor equation for the solution (\ref{geom1}). This will allow us to correctly identify left and right-moving symmetries. We adopt our supersymmetry conditions from \cite{Kelekci:2014ima}. We introduce a natural orthonormal frame: 
\bea
e^{\rho} &=& \dd \rho, \quad e^{+} = \frac{1}{\sqrt{2}} e^{\rho} \dd \gamma, \quad e^{-} = \frac{1}{\sqrt{2}} e^{\rho} \dd \bar{\gamma}, \nn
e^{\theta} &=& \frac{1}{2} \dd \theta, \quad e^{\phi} = \frac{1}{2} \sin \theta \dd \phi, \quad e^{\psi} = \frac{1}{2} ( \dd \psi + \cos \theta \dd \phi). 
\eea

From the dilatino variation, we find the projector: 
\be
\Gamma^{\rho + - \theta \phi \psi} \epsilon_{\pm} = \epsilon_{\pm}. 
\ee

Solving for $\epsilon_{\pm}$ one gets: 
\bea
\epsilon_{+} = e^{\frac{\rho}{2} \Gamma^{+-}} e^{\frac{\bar{\gamma}}{\sqrt{2}} \Gamma^{\rho +}} e^{-\frac{\psi}{2} \Gamma^{\theta \phi} } \tilde{\epsilon}_{+}, \quad \epsilon_{-} = e^{-\frac{\rho}{2} \Gamma^{+-}} e^{\frac{\gamma}{\sqrt{2}} \Gamma^{\rho -}} e^{\frac{\theta}{2} \Gamma^{\phi \psi } } e^{\frac{\phi}{2} \Gamma^{\theta \phi} } \tilde{\epsilon}_{-}. 
\eea
Here we observe a clean split between the spinors describing the left and right supersymmetries, where $\epsilon_{+}$ depends on $\psi$ and $\epsilon_{-}$ depends on $(\theta, \phi)$. We see that T-duality on $\psi$ breaks all supersymmetries associated to $\epsilon_+$. The shift in $\gamma$ then preserves the $\epsilon_{-}$ spinors satisfying $\Gamma^- \tilde{\epsilon}_- = 0$. This appears to leave $\mathcal{N} = (2,0)$ supersymmetry, but we can confirm this statement by having a quick look at the Killing spinor equations of the \textit{deformed} geometry. 

To this end, let us reconsider the orthonormal frame: 
\bea
e^{\rho} &=& \dd \rho, \quad e^{+} = \frac{1}{\sqrt{2}} e^{\rho} ( \dd \gamma + A), \quad e^{-} = \frac{1}{\sqrt{2}} e^{\rho} \dd \bar{\gamma}, \nn
e^{\theta} &=& \frac{1}{2} \dd \theta, \quad e^{\phi} = \frac{1}{2} \sin \theta \dd \phi, \quad e^{\psi} = \frac{1}{2} ( \dd \psi + \cos \theta \dd \phi)
\eea
where $A$ is now a one-form on $S^3$. The ansatz for $B$ is 
\be
B = \frac{1}{2} e^{2 \rho} ( \dd \gamma + A) \wedge \dd \bar{\gamma} + \frac{1}{4} \cos \theta \dd \phi \wedge \dd \psi. 
\ee

From the dilatino variation for the deformed solution, we find an additional projector $\Gamma^{-} \tilde{\epsilon}_{\pm} = 0$, so supersymmetry is broken to $\mathcal{N} = (2,2)$. From the $\delta_{\bar{\gamma}} \psi$ gravitino variation, we find the following equation: 
\be
\delta_{\bar{\gamma}} \psi = \frac{1}{\sqrt{2}} e^{- \rho} \partial_{\bar{\gamma}} \epsilon - \frac{1}{2} \Gamma^{\rho + } \sigma_3 \epsilon - \frac{1}{2} \Gamma^{\rho +} \epsilon - \frac{1}{8 \sqrt{2}}e^{\rho} F_{ab} \Gamma^{ab } \epsilon - \frac{1}{8 \sqrt{2}} e^{\rho} F_{ab} \Gamma^{ab} \sigma_3 \epsilon = 0, 
\ee
where we have introduced $F = \dd A$. Note that due to the $\rho$ dependence of the terms is different, so the only way we can solve this equation is if $\tilde{\epsilon}_+ =0$. This is in line with expectations, since the Killing spinor dependence of $\psi$ should vanish. Thus, the remaining supersymmetry is $\mathcal{N} = (2,0)$, as claimed.

\end{document}